\documentclass[11pt]{article}
\usepackage{moriond,epsfig}

\def\Journal#1#2#3#4{{#1} {\bf #2}, #3 (#4)}

\def\be{\begin{equation}}
\def\ee{\end{equation}}
\def\bea{\begin{eqnarray}}
\def\eea{\end{eqnarray}}

\begin{document}
\vspace*{4cm}
\title{BINGO: a single dish approach to 21cm intensity mapping}

\author{R.A. Battye$^{1}$, M.L. Brown$^{1}$, I.W.A. Browne$^{1}$,  R.J. Davis$^{1}$, P. Dewdney$^{2}$, C. Dickinson$^{1}$, \\ G. Heron$^{1}$, B. Maffei$^{1}$, A. Pourtsidou$^{1,3}$, P.N. Wilkinson$^{1}$}

\address{$^{1}$Jodrell Bank Centre for Astrophysics, University of Manchester, Oxford Road, Manchester, M13 9PL.\\
$^{2}$ SKA Organization, Alan Turing Building, Manchester,  M13 9PL. \\
$^{3}$Dipartimento di Astronomia, Universit\`a di Bologna, via Ranzani 1,
 I-40127 Bologna, Italy.
}

\maketitle\abstracts{BINGO is a concept for performing a 21cm intensity mapping survey using a single dish telescope. We briefly discuss the idea of intensity mapping and go on to define our single dish concept. This involves a $\sim 40\,{\rm m}$ dish with an array of $\sim 50$ feed horns placed $\sim 90\,{\rm m}$ above the dish using a pseudo-correlation detection system based on room temperature LNAs and one of the celestial poles as references. We discuss how such an array operating between 960 and $1260\,{\rm MHz}$ could be used to measure the acoustic scale to $2.4\%$ over the redshift range $0.13<z<0.48$ in around 1 year of on-source integration time by performing a $10\,{\rm deg}\times 200\,{\rm deg}$ drift scan survey with a resolution of $\sim 2/3\,{\rm deg}$. }

\section{21cm intensity mapping}
 
The studies of dark energy~\cite{Copeland:2006} and modified gravity~\cite{Clifton:2011} are now a major area of research using a plethora of observational tests~\cite{Weinberg:2012}. The baryonic acoustic oscillation (BAO) approach~\cite{Eisenstein:2003} is now well established. Originally applied to the SDSS LRG DR3 survey~\cite{Eisenstein:2005} subsequent detections have been reported by a number of groups analysing data from 2dF~\cite{Cole:2005}, SDSS DR7~\cite{Percival:2010}, WiggleZ~\cite{Blake:2011}, 6dF~\cite{Beutler:2011}and BOSS~\cite{Anderson:2012} surveys. These surveys detect galaxies in the optical and construct a galaxy density contrast whose 2-point correlation function, or equivalent power spectrum, is then estimated. Under the assumption of simple biasing between the galaxy density field and that of the underlying matter density, one can infer the acoustic scale.

Galaxy redshifts obtained using the 21cm line of neutral hydrogen can also be used to perform redshift surveys~\cite{Abdalla:2005,Duffy:2008} but this typically requires very large collecting areas to do this at high redshift - this was the original motivation for the SKA~\cite{Wilkinson:1991}.  Radio telescopes with apertures $\sim 100\,{\rm m}$ have sufficient surface brightness sensitivity to detect HI at high redshift if they have a filled or close to filled aperture, but they will preferentially detect objects of angular size comparable to their resolution~\cite{Battye:2004} which will typically be clusters. The idea of 21cm intensity mapping~\cite{Peterson:2006,LoebWyithe:2008,Chang:2008,Pen:2008,Chang:2010,Peterson:2012} is to use the full intensity field $T(f,\theta,\phi)$ and measure the power spectrum directly, without ever detecting the individual galaxies. The average signal is expected to be $\sim 100\,\mu{\rm K}$ on degree scales with the band width of $\sim\,\hbox{few}\,{\rm MHz}$, with order one fluctuations. This will require exquisite subtraction of receiver bandpasses and the continuum emission from our own Galaxy and extragalactic sources. The latter of these two should be possible using the fact that the continuum emission should have a relatively smooth spectral signature relative the 21cm emission and designing a telescope to mitigate the former is the main topic of this paper.

\section{BINGO concept} 

A detailed analysis of the signal and possible observing strategies~\cite{Battye:2012at} suggests that a survey of $2000\,{\rm deg}^2$ with a resolution of $2/3\,{\rm deg}$ in the frequency band $960-1260\,{\rm MHz}$ is close to optimal in terms of measurement of the acoustic scale. We propose a static single dish telescope to achieve this by a drift scan survey at constant declination. A key issue is the size of the smallest dimension, $\theta_{\rm min}$, of the survey area since power spectrum, $P(k)$ needs to be binned with $\Delta k>\pi/\theta_{\rm min}r(z)$ where $r(z)$ is the coordinate distance to redshift $z$, so that the bandpowers are uncorrelated, but this bin width needs to be a factor of around 3 times the acoustic scale, $k_{\rm A}\approx (150\,{\rm Mpc})^{-1}$ since otherwise the BAOs would be washed out. This restricts $\theta_{\rm min}>10\,{\rm deg}$ for $z\approx 0.3$. If the drift scan is at  moderate latitudes then the largest  dimension of the survey will be $\approx 200\,{\rm deg}$ and hence a survey of $\approx 2000\,{\rm deg}^2$ will be possible with around 50 feedhorns in 3 rows giving an instantaneous field-of-view of $10\,{\rm deg}\times 2\,{\rm deg}$.

\begin{figure}
\begin{center}
\includegraphics[height=9.0cm,width=4.5cm]{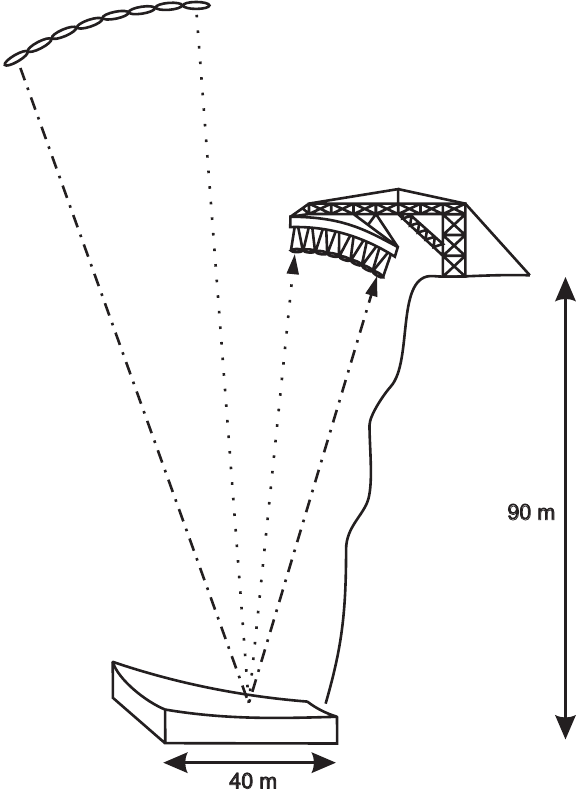}\hskip 1.0cm
\includegraphics[height=5.0cm,width=10.0cm]{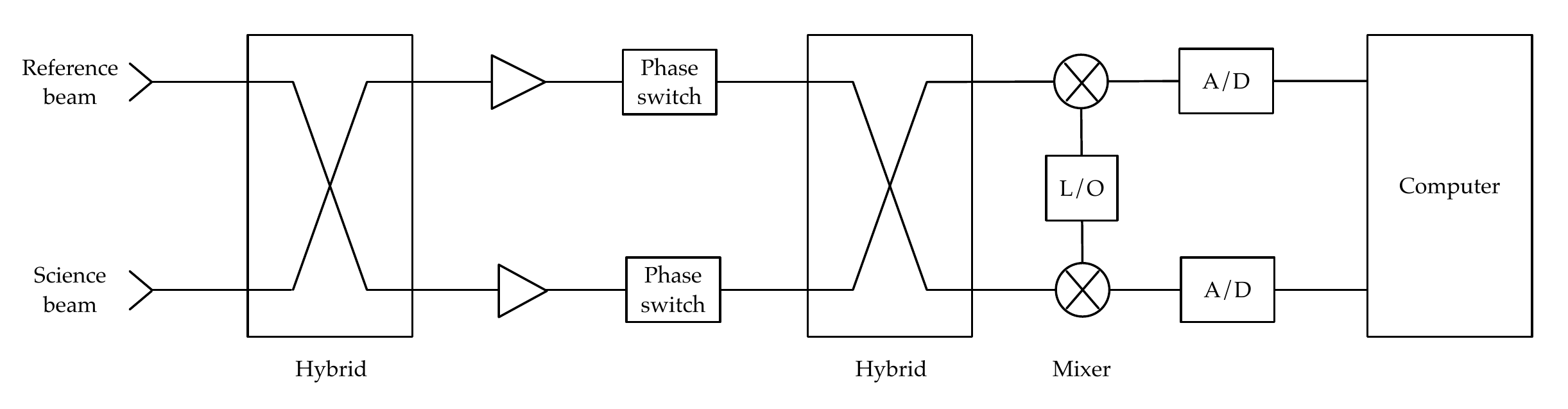}
\caption{On the left schematic of the proposed design of the BINGO telescope. There will be an under-illuminated $\sim 40\,{\rm m}$ static parabolic reflector at the bottom of a cliff which is around $\sim  90\,{\rm m}$ high. A boom will be placed at the top of a cliff  on which there is a  receiver system of $\sim 50$ feed-horns. On the right a block diagram for the receiver chain for the proposed pseudo-correlation receiver system. The reference beam will point toward one of the celestial poles.}
\label{fig:telescope}
\end{center}
\end{figure}

In order to achieve the desired resolution the feedhorns will need to illuminate an aperture of $\approx 25\,{\rm m}$ and the total dish size is proposed to be $\approx 40\,{\rm m}$ in order to reduce the sidelobe levels. It would be difficult to accommodate such a large feedhorn array in a standard focal arrangement since the focal area will be too small. Therefore, we propose to suspend the feedhorns $>90\,{\rm m}$ above the dish on a boom  which is anchored to a cliff, or other steep slope. The basic arrangement is illustrated schematically in the in the left-hand panel of Fig.~\ref{fig:telescope}. This should allow a focal plane sufficiently large to allow $\theta_{\rm min}>10\,{\rm deg}$. The parameters of the proposed system are presented in table~\ref{tab:parameters}

We will use a pseudo-correlation receiver system in order to control $1/f$ fluctuations~\cite{Rohlfs:2004}. A block diagram for such a system is presented in the right-hand panel of  Fig.~\ref{fig:telescope}. It uses hybrids to difference and add signals either side of the primary amplification. If the hybrids are perfect and the inputs are perfectly balanced then the $1/f$ noise due to gain fluctuations is completely removed in the difference spectrum. The receiver system will operate between 960 and $1260\,{\rm MHz}$ in order to avoid the mobile phone  band between 900 and $960\,{\rm MHz}$ and will use room temperature LNAs with a conservative system temperature of $T_{\rm sys}=50\,{\rm K}$. The reference beam of each pseudo-correlation system will be pointed at one of the celestial poles without going via the main dish in order to provide a stable, low resolution reference with the same spectrum as the sky.

\begin{table}
 \caption{Summary of proposed BINGO telescope parameters}
 \vspace{0.4cm}
 \begin{center}
 \begin{tabular}{|c|c|}
 \hline
Main reflector diameter & 40~m \\
Illuminated diameter & 25~m    \\
Resolution (at $30\,{\rm cm}$) & $\sim$40 arcmin   \\
Number of feeds & 50     \\
Instantaneous field of view & 10\,${\rm deg}\times 2\,{\rm deg}$ \\
Frequency range & $960\,{\rm MHz}$ to $1260\,{\rm MHz}$ \\
Number of frequency channels & $\geq$300  \\
 \hline
 \end{tabular}
 \end{center}
 \label{tab:parameters}
 \end{table}

\section{Projected science reach}

\begin{figure}
\centering
\includegraphics[height=7.0cm,width=8.0cm]{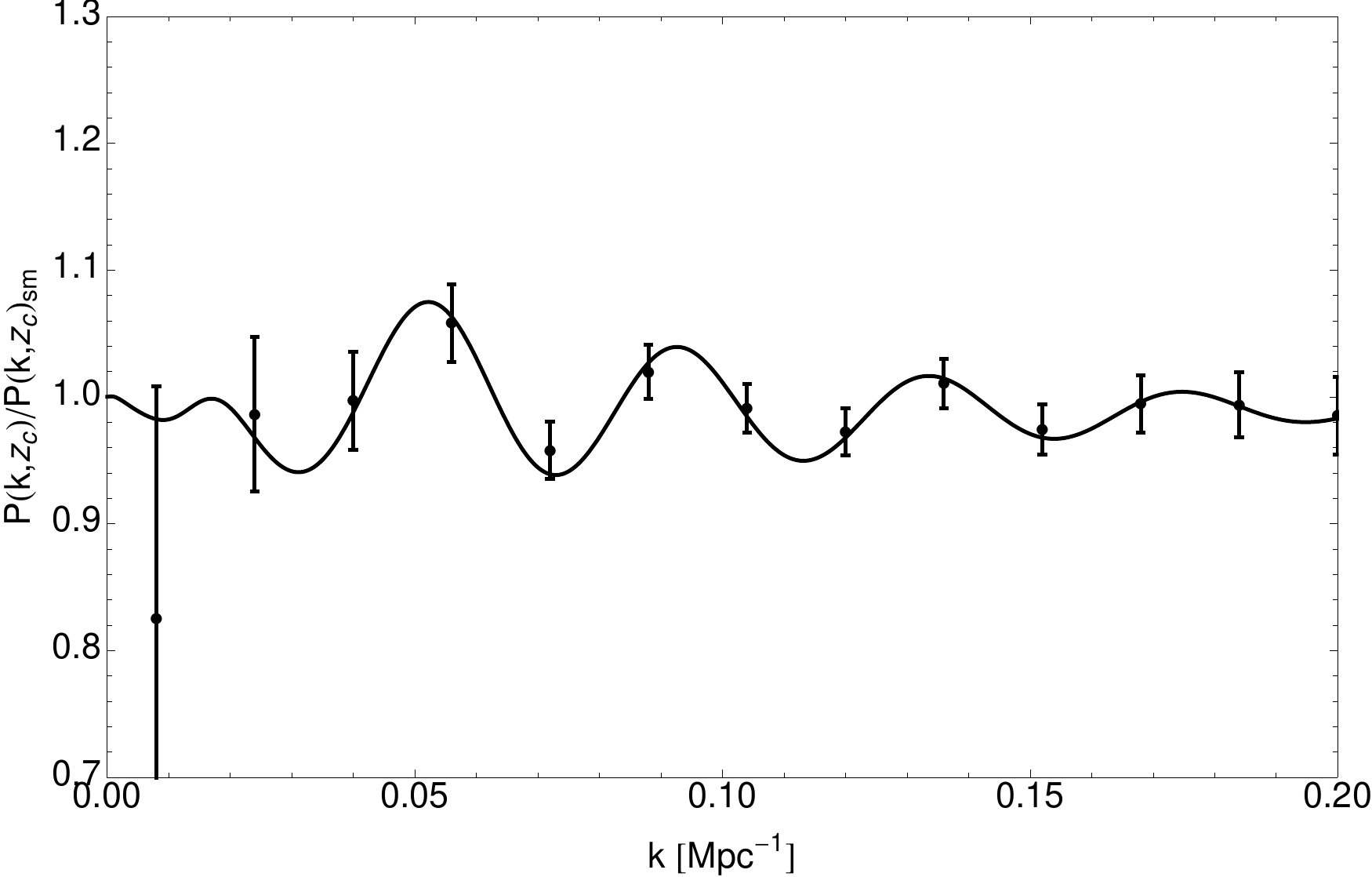}\includegraphics[height=7.0cm,width=8.0cm]{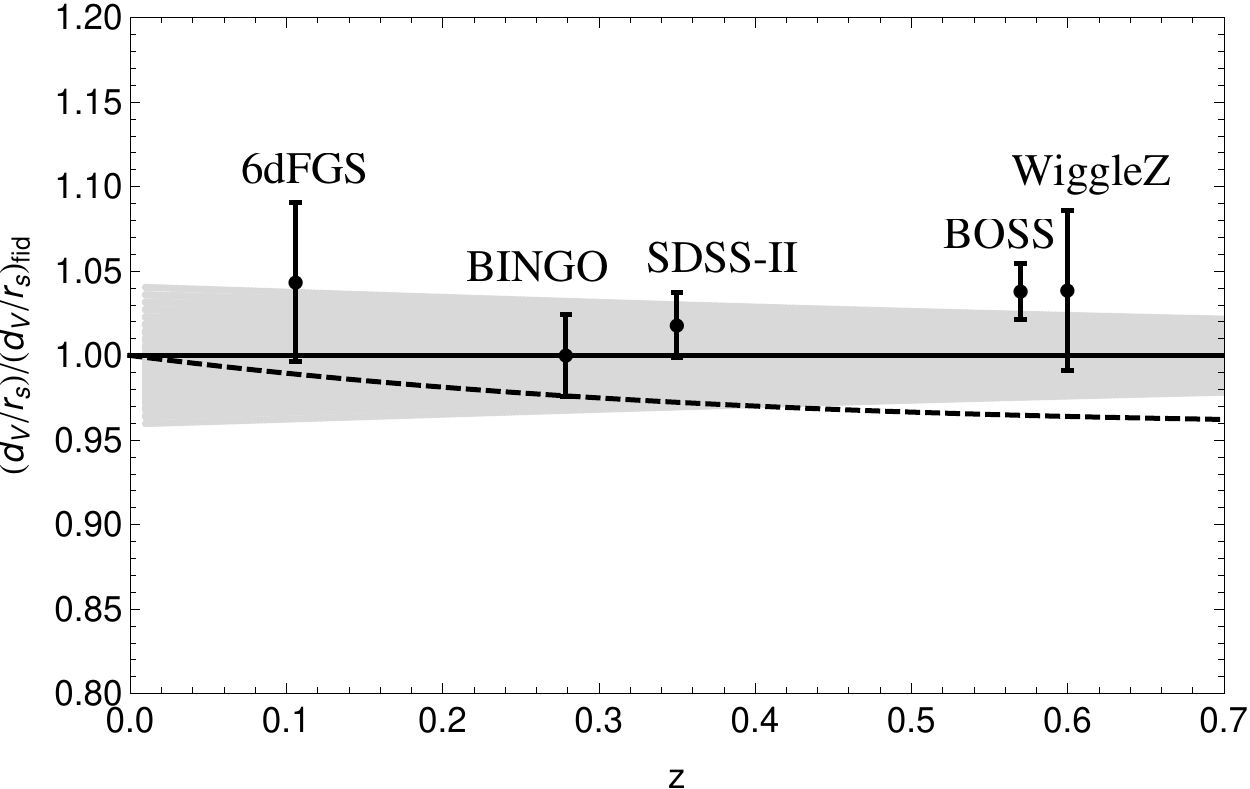}
\caption{On the left projected errors on the power spectrum (divided by a smooth power spectrum) expected for the survey described in the text. We have used $\Delta k = 0.016\,{\rm Mpc}^{-1}$. The projected errors would lead to a measurement of the acoustic scale with a percentage fractional error of $2.4\%$. On the right, projected constraints on the residual Hubble diagram for the volume averaged distance, $d_{\rm V}(z)$ from a fiducial model. Included also are the actual measurements made by 6dF, SDSS-II, BOSS and WiggleZ. The shaded region represents indicates the range of $d_{\rm V}$ allowed by the $1\sigma$ constraint $\Omega_{\rm m}h^2$ from WMAP7. The dotted line is the prediction for $w=-0.84$.}
\label{fig:science}
\end{figure}

The projected error on a power spectrum measurement average over a radial bin in $k$-space of width $\Delta k$ is 
\begin{equation}
{\sigma_P\over P}=\sqrt{2{(2\pi)^3\over V_{\rm sur}}{1\over 4\pi k^2\Delta k}}\left(1+{\sigma_{\rm pix}^2V_{\rm pix}\over [{\bar T}(z)]^2W(k)^2P}\right)\,,
\end{equation}
where $V_{\rm sur}$ is the survey volume, $\sigma_{\rm pix}$ is the pixel noise over a nominal bandwidth of $\Delta f=1\,{\rm MHz}$, $W(k)$ is the angular window function and ${\bar T}(z)$ is the average temperature  
\begin{equation}
{\bar T}(z)=44\,\mu{\rm K}\left({\Omega_{\rm HI}(z) h\over 2.45\times 10^{-4}}\right){(1+z)^2\over E(z)}\,.
\end{equation}
$\Omega_{\rm HI}$ is the HI density relative to the present day critical density and $E(z)=H(z)/H_0$.  We have computed the projected errors on the measurement of the power spectrum at $z=0.28$ for $1\,{\rm year}$ of on-source integration (which we would expect to be possible in around 2 years of observing), corresponding to the central redshift of the proposed survey, and they are presented on the power spectrum relative to a smooth spectrum in the left-hand panel of Fig.~\ref{fig:science}. It is clear from this that it should be possible to measure the acoustic scale. We note that these estimates ignore the possible effects of foreground removal.

Such a measurement will allow the acoustic scale to be measured to $\delta k_{\rm A}/k_{\rm A}\approx 0.024$. The science reach of such a measurement  is illustrated in the right-hand panel of Fig.~\ref{fig:science} where it is compared to the recent measurements. It is clear that this would lead to interesting constraints on the cosmological parameters. Assuming that the dark energy equation  of state parameter, $w=P/\rho$, is constant  and all the other cosmological parameters are constant, this would lead to $\delta w/w\approx 0.16$.

\end{document}